\documentclass[10pt,thmsa]{article}
\usepackage{amssymb}

\input{tcilatex}
\begin{document}

\title{Modular Wedge Localization and the d=1+1 Formfactor Program }
\author{\textbf{Bert Schroer} \\
%EndAName
Freie Universit\"{a}t Berlin,\\
Institut f\"ur Theoretische Physik}
\maketitle

\begin{abstract}
In this paper I continue the study of the new framework of modular
localization and its constructive use in the nonperturbative d=1+1
Karowski-Weisz-Smirnov formfactor program. Particular attention is focussed
on the existence of semilocal generators of the wedge-localized algebra
without vauum polarization (FWG-operators) which are closely related to
objects fulfilling the Zamolodchikov-Faddeev algebraic structure. They
generate a ``thermal Hilbert space'' and allow to understand the equivalence
of the KMS conditions with the so-called cyclicity equation for formfactors
which was known to be closely related to crossing symmetry properties. The
modular setting gives rise to interesting new ideas on ``free'' d=2+1 anyons
and plektons.
\end{abstract}

\section{Introduction}

The historical roots of ``Modular Localization'' as a kind of inversion of
the famous Bisognano-Wichmann observation \cite{Bi Wi} and its relation with
the mathematical Tomita-Takesaki modular theory \cite{Takesaki} of von
Neumann algebras in ``general position'' and their thermal KMS structure 
\cite{Haag} as well as their constructive power for low-dimensional QFT in
general and the formfactor problem in particular, has already been
highlighted in previous work of the author \cite{S1}. In addition there is a
recent paper \cite{Nieder} in which, similar to the present work, the
thermal aspect was applied to a specific problem arising in the formfactor
problem.

My original interest in this subject arose from the attempt to understand
the loss of unicity when one passes from the unique $(m,s)$-Wigner
representation to free fields: there are as many free fields in the $(m,s)$%
-Fock space as there are intertwiners between the $D^{(s)}(R(\Lambda
,\Lambda ^{-1}p))$ Wigner rotation matrices and the $D^{\left[
n_{+},n_{-}\right] }(\Lambda )$ finite dimensional representation matrices
of the Lorentz group \cite{Wei}. All of these infinitely many fields \textit{%
describe the same spacetime physics,} even though most of them are not
``Eulerian''\footnote{%
Meaning that, similar as for the Dirac equation, the tranformation law of
the field follows from the structure of the differential equation. The
Wigner representation theory was, among other things, introduced just to
avoid the problem of ambiguities in equations of motions and their
proliferation.} or Lagrangian. The net of local algebras associated with the 
$(m,s)$-Wigner representation is again unique, and the method of its direct
construction (bypassing the free fields) is based on modular localization.
Although this has been described in the cited literature, we repeat the main
arguments in the next section in order to keep this paper reasonably
self-contained. The new localization modular applied to wave functions is
more suitable for QFT than the old one of Newton and Wigner \cite{Haag}, who
adapted the Born (probability) localization of Schr\"{o}dinger theory to the
Wigner representation spaces. In a similar vein as the existence of the
Klein paradox, the difference in localization shows the limitations of
analogies between wave functions of relativistic particles and
Schr\"{o}dinger wave functions. Modular localization in the context of the
Wigner theory requires us to interpret the complex representation space as a
``twice as large'' real Hilbert space with an appropriately defined real
orthogonal (or equivalently symplectic) inner product. The real
orthogonality physically means causally ``opposite'' in the sense of quantum
theoretical localization and this concept in Fock space passes directly to
the von Neumann notion of compatibility of measurements in the sense of
commutants of von Neumann algebras.

The third section is a continuation of the second section on modular aspects
of ``Wignerism''; but this time with a stronger emphasis on thermal KMS
(Hawking, Unruh) properties of wedge localization and crossing symmetry.

In the context of interacting theories one obtains a framework which is
totally intrinsic and characteristic of local quantum physics (LQP).
Concepts which are shared with nonrelativistic quantum physics and play a
crucial role in perturbation theory, as the interaction picture,
time-ordering, euclidean path integrals etc. \textit{are not required} and
unnatural in this approach. No-Go theorems, as the Haag theorem (an
obstruction against the validity of the field theoretic interaction
picture), are not valid for analogous modular concepts, since one has
unitary equivalence between the wedge-restricted incoming (free) and
interacting algebras. In the terminology explained in section 4 there is no
Haag-theorem obstruction against the existence of the ``modular M\o ller
operator''.

There is of course one construction in the literature which \textit{also
does not use} the above \textit{standard} text book \textit{formalism: the
bootstrap formfactor approach} to factorizable d=1+1 models of Karowski and
Weisz \cite{KW} as axiomatized and extended by Smirnov \cite{Smi}. This has
generated the erroneous impression with some physicist that there are two
different subjects: standard textbook kind of QFT, and 2-dimensional
factorizable- (as well as chiral conformal-) QFT. This paper is also
intended to counteract such impressions.

No relativistic QFT allows a physically consistent modifications in the
direction of momentum space cutoffs or regularizations. In particular it is
not possible to manufacture a factorizing model with such a modification
without wrecking its physical interpretation altogether. If one wants to
have versions outside relativistic quantum physics, but with a similar
particle interpretation and scattering theory (without Einstein causality),
one has to find a lattice version. This is a highly nontrivial task in its
own right; the derivation of cluster properties (and hence Haag-Ruelle)
scattering theory without Einstein causality is much more difficult \cite{Ba
Fr}. In particular our modular localization concepts would not apply to
lattice theories. Our methods in this article are totally characteristic of
local quantum physics; they are wrecked by cutoffs, regulators and
elementary lengths.

The main part of this paper is the fourth section where we relate modular
localization with factorizing models. A useful illustration (despite its
physical austerity) is supplied by the Federbush model \cite{Wigh}\cite{SW}.
Its explicitly known solution is similar to the Ising model QFT \cite
{Karowski}, but its rapidity independent S-matrix depends continuously on a
coupling strength. The reason why it did not hitherto appear in the
factorizing list is that the S-matrix bootstrap was restricted to parity
conserving theories. The main message from this model and the principles of
algebraic QFT is that the simplicity of factorizable models does not so much
show up in the generalized formfactors of ``field coordinates'' and
correlation functions (which apart from the Ising theory and the Federbush
model mostly remain too complicated for explicit analytic calculations), but
rather in the computation of the (net of) wedge algebras (and the ensuing
double cone algebras). The net of wedge algebras uniquely fixes the total
net, and since the wedge generators turn out to be determined uniquely by
the factorizing S-matrix, the so called inverse problem of QFT has a unique
solution in the context of factorizing models.  These algebras and their
mutual relations are sufficient for exploring and characterizing the
intrinsic physical content of a real time QFT\footnote{%
The constructive aspect of the formfacor approach to renormalizable
factorizing models and the present new modular method reveals more structure
than the old method of constructive QFT with respect to e.g. $\phi _{2}^{4}$
which was modelled on Lagrangian perturbation theory. The new construction
is not impeded by worse than canonical short distance behaviour. In fact the
issue of short distance behavior beyond the limits set by causality does not
enter the nonperturbative modular construction.}, since the S-matrix and
perhaps some distinguished Noether current fields (energy momentum tensor,
charge currents) are not only sufficient mathematical characteristics, but
are also the only experimentally accessible observables in high energy
physics. In statistical mechanics also the individual correlation functions
of certain order/disorder fields are measurable and for those one has to
work harder than for the structure of the algebras of real time QFT.
Fortunately the computation of e.g. the spectrum of critical indices is part
of the much simpler real time spin-statistics (the statistical phases \cite
{FRSII}) discussion of algebraic QFT.

The concluding remarks offers a scenario for higher dimensional modular
constructions in the presence of interactions. In that case one does not
have an S-matrix bootstrap program which could be \textit{separated from the
rest of QFT}. Although algebraic QFT offers some hints, one its still very
far from covering these ideas with an analytically accessible formalism.

Mathematically the problems at hand are a special case of the so called 
\textit{inverse problem of modular theory} i.e. the reconstruction of a von
Neumann algebra from its modular data \cite{Wollenberg}. Only in its more
physical setting where it passes to the \textit{inverse problem of QFT},
namely the question whether \textit{physically admissable scattering
matrices possess a unique local QFT}, the uniqueness can be established
under reasonable physical assumptions. This will be shown in a subsequent
paper \cite{SWi}. For a mathematical and conceptual background I refer to 
\cite{schroer} and to my notes \cite{Notes} which are intended as part of a
planned monograph on ``Nonperturbative Approach to Local Quantum Physics''.
There the reader also finds (in chapter 2) an account of the unpublished
work \cite{Ley} adapted to the present setting.

\section{Liberation from Free Field Coordinates}

As explained elsewhere \cite{S1}, one may use the Wigner representation
theory for positive energy representations in order to construct fields from
particle states. For $d=3+1$ space-time dimensions there are two families of
representation: $(m,s)$ and $(0,h)$. Here $m$ is the mass and designates
massive representation and $s$ and $h$ are the spins resp. the helicities $h$%
. These are invariants of the representations (``Casimirs'') which refer to
the Wigner ``little'' group; in the first case to SU(2) in which case $s=$
(half) integer, and for $m=0$ to the little group (fixed point group of a
momentum $\neq 0$ on the light cone) $\tilde{E}(2)$ which is the two-fold
covering of the euclidean group in the plane. The zero mass representations
in turn split into two families. For the ``neutrino-photon family'' the
little group has a non-faithful representation (the ``translative'' part is
trivially represented) whereas for Wigner's ``continuous $h$
representation'' the representation is faithful and infinite component, but
allows no identification with known zero mass particles.

In the massive case, the transition to covariant fields is most conveniently
done with the help of intertwiners between the Wigner spin $s$
representations $D^{(s)}(R(\Lambda ,p))$ which involve the $\Lambda ,p$
dependent Wigner rotation $R$ and the finite dimensional covariant
representation of the Lorentz-group $D^{[A,B]}$ 
\begin{equation}
u(p)D^{(s)}(R(\Lambda ,p))=D^{[A,B]}(\Lambda )u(\Lambda ^{-1}p)  \label{1}
\end{equation}
The only restriction is: 
\begin{equation}
\mid A-B\mid \le s\le A+B  \label{2}
\end{equation}
which leaves infinitely many $A,B$ (half integer) choices for a given $s$.
Here the $u(p)$ intertwiner is a rectangular matrix consisting of $2s+1$
column vectors $u(p,s_{3}),s_{3}=-s,...,+s$ of length $(2A+1)(2B+1)$. Its
explicit construction using Clebsch-Gordan methods can be found in
Weinberg's book \cite{Wei}. Analogously there exist antiparticle (opposite
charge) $v(p)$ intertwiners: $D^{(s)*}(R(\Lambda ,p)\longrightarrow
D^{[A,B]}(\Lambda )$. The covariant field is then of the form: 
\begin{eqnarray}
\psi ^{[A,B]}(x) &=&\frac{1}{(2\pi )^{3/2}}\int
(e^{-ipx}\sum_{s_{3}}u(p_{1},s_{3})a(p_{1},s_{3})+  \label{field} \\
&&+e^{ipx}\sum_{s_{s}}v(p_{1},s_{3})b^{*}(p_{1},s_{3}))\frac{d^{3}p}{2\omega 
}  \nonumber
\end{eqnarray}
where $a(p)$ and $b^{*}(p)$ are annihilation (creation) operators in a Fock
space for particles (antiparticles). The bad news (only at first sight, as
it fortunately turns out) is that we lost the Wigner unicity: there are now
infinitely many $\psi ^{[A,B]}$ fields with varying $A,B$ but all belonging
to the same $(m,s)$-Wigner representation and living in the same Fock space.
Only one of these fields is ``Eulerian'' (examples: for $s=\frac{1}{2}$
Dirac, for $s=\frac{3}{2}$ Rarita-Schwinger) i.e. the transformation
property of $\psi $ is a consequence of the nature of a linear field
equation which is derivable by an action principle from a Lagrangian.
Non-Eulerian fields as e.g. Weinberg's $D^{\left[ j,0\right] }+D^{\left[
0,j\right] }$ fields for $j\geq \frac{3}{2},$ cannot be used in a canonical
quantization scheme or in a formalism of functional integration because the
corresponding field equations have more solutions than allowed by the
physical degrees of freedom ( in fact they have tachyonic solutions). The
use of formula (\ref{field}) with the correct $u,v$ intertwiners in the (on
shell) Bogoliubov-Shirkov approach (which different from off shell
approaches as euclidean functional integrals) does not require the existence
of free bilinear Lagrangians) based on causality is however legitimate.
Naturally from the point of view of the Wigner theory, which is totally
intrinsic and does not use quantization ideas, there is no preference of
Eulerian versus non Eulerian fields.

It turns out that the above family of fields corresponding to $(m,s)$
constitute the linear part of the associated ``Borchers class'' \cite{Haag}.
For bosonic fields the latter is defined as: 
\begin{equation}
B(\psi )=\left\{ \chi (s)\mid [\chi (x),\psi (y)]=0,\,\,\,(x-y)^{2}<0\right\}
\end{equation}
If we only consider cyclic (with respect to the vacuum) relatively local
fields, then we obtain transitivity of the causality for the resulting
fields. This class depends only on $\left( m,s\right) $ and is generated by
the Wick-monomials of $\psi $. A mathematically and conceptually more
manageable object which is manifestly independent of the chosen $(m,s)$
Fock-space field, is the local von Neumann algebra $\mathcal{A}(\mathcal{O})$
generated by $\psi $: 
\begin{equation}
\mathcal{O}\rightarrow \mathcal{A}(\mathcal{O},\psi )=\mathcal{A}(\mathcal{O}%
,\chi )
\end{equation}
Here $\chi \sim \psi $ is any cyclic (locally equivalent) field in the same
Borchers class of $\psi .$

Now we have reached our first goal: the lack of uniqueness of local $(m,s)$
fields is explained in terms of the arbitrariness in the choice of ``field
coordinates'' which generate the same net of von Neumann algebras. According
to the physical interpretation in algebraic QFT this means that the physics
does not depend on the concretely chosen (cyclic) field.

Since algebraic QFT shuns inventions and favors discoveries, it is deeply
satisfying that there are arguments to the extend that every causal net
fulfilling certain spectral properties is automatically ``coordinatizable''.
For chiral conformal theories there exists even a rigorous proof \cite{Joe}.
So one can be confident that the physical content has not been changed as
compared to the standard Wightman approach which in turn was abstracted from
a synthesis of the Heisenberg-Pauli canonical quantization of classical
Lagrangian field theory with the Wigner representation theory of the
Poincar\'{e} group. The use of local field coordinates tends to make
geometric localization properties of the algebras manifest. But only if
there exist pointlike covariant generators which create charged states
(counter example: for Maxwellian charges they do not exist; the physical
electron field is noncompactly localized) the localization can be encoded
into classical smearing function. The localization concept is ``maximally
classical'' for the free Weyl and CAR algebras which in fact are just
function algebras with a noncommutative product structure. For these special
cases the differential geometric concepts as fibre bundles may be directly
used in local quantum physics. Outside of these special context where
quantum localization can be described in terms of support properties of
classical functions, the only sufficiently general reliable concepts are
those based on von Neumann algebra methods of algebraic QFT. In that case
the quantum localization may deviate from the classical (differential-)
geometric concepts \cite{schroer} and the use of ``field coordinates'' is
less useful than that of local nets. This however does not mean that
geometrical concepts are useless, but rather that they cannot be imposed but
must be derived as a consequence of local quantum physics.

In the following we describe a way to construct the interaction-free nets
directly \cite{S1}, thus bypassing the use of field coordinates altogether.
We use the d=3+1 Wigner $(m,s)$-representations as an illustrative example.
In case of charged particles (particles$\neq $antiparticles) we double the
Wigner representation space: 
\begin{equation}
H=H_{Wig}^{p}\oplus H_{Wig}^{\bar{p}}
\end{equation}
in order to incorporate the charge conjugation operation as an (antilinear
in the Wigner theory) operator involving the p-\={p}-flip. On this extended
Wigner space one can act with the full Poincar\'{e} group (where those
reflections which change the direction of time are antiunitarily
represented). For the modular localization in a wedge we only need the
standard L-boost $\Lambda (\chi )$ and the standard reflection $r$ which (by
definition) are associated with the $t-x$ wedge: 
\begin{equation}
\delta ^{i\tau }\equiv \pi _{Wig}(\Lambda (\chi =2\pi \tau ))
\end{equation}
\begin{equation}
j\equiv \pi _{Wig}(r)
\end{equation}
These operators have a simple action on the p-space (possibly) doubled
Wigner wave functions, in particular: 
\begin{equation}
(j\psi )(p)\simeq \left( 
\begin{array}{ll}
0 & -1 \\ 
1 & 0
\end{array}
\right) \bar{\psi}(p_{0},p_{1},-p_{2},-p_{3})
\end{equation}
By functional calculus we form $\delta ^{\frac{1}{2}}$ and define: 
\begin{equation}
s\equiv j\delta ^{\frac{1}{2}}
\end{equation}
This unbounded antilinear densely defined operator $s$ is involutive on its
domain: $s^{2}=1.$ Its -1 eigenspace\footnote{%
It does not matter whether we take the + or - sign for the characterization
of modular localization since we can convert one into the other via
multiplication with $i$ and the dense domains $H_{R}(W)+iH_{R}(W)$ are the
same. This is one of the rare occasions where a sign error remais without
serious consequences.} is a real closed subspace $H_{R}$ of $H$ which allows
the following characterization of the domain of $s:$%
\begin{eqnarray}
dom(s) &=&H_{R}+iH_{R} \\
s(h_{1}+ih_{2}) &=&-h_{1}+ih_{2}  \nonumber
\end{eqnarray}
Defining: 
\begin{equation}
H_{R}(W)\equiv U(g)H_{R},\,\,\,W=gW_{stand}
\end{equation}
where g is an appropriate Poincar\'{e} transformation, we find the following
theorem (D. Guido, private communication):

\begin{theorem}
$H_{R}(W)$ is an isotonous net of real Hilbert spaces i.e. $%
H_{R}(W_{1})\subsetneq H_{R}(W_{2})$ if $W_{1}\subsetneq W_{2}$.
\end{theorem}

Its proof, which follows from a theorem by Borchers \cite{Bo} will nor be
given here

If we now define: 
\begin{equation}
H_{R}(\mathcal{O})\equiv \bigcap_{W\supset \mathcal{O}}H_{R}(W)
\end{equation}
then it is easily seen (even without the use of the u,v-intertwiners) that
the spaces $H_{R}(\mathcal{O})+iH_{R}(\mathcal{O})$ are still dense in $%
H_{Wig}$ and that the formula: 
\begin{equation}
s(\mathcal{O})(h_{1}+ih_{2})\equiv -h_{1}+ih_{2}
\end{equation}
defines a closed involutive operator with a polar decomposition: 
\begin{equation}
s(\mathcal{O})=j(\mathcal{O})\delta (\mathcal{O})^{\frac{1}{2}}
\end{equation}
Although now $j(O)$ and $\delta (\mathcal{O})^{i\tau }$ have no obvious
geometric interpretation, there is still a bit of geometry left, as the
following theorem shows:

\begin{theorem}
The $H_{R}(\mathcal{O})$ form an orthocomplemented net of closed real
Hilbert spaces, i.e. the following ''duality'' holds: 
\begin{equation}
H_{R}(\mathcal{O}^{\prime })=H_{R}(\mathcal{O})^{\prime }=iH_{R}^{\bot }(%
\mathcal{O}).
\end{equation}
\end{theorem}

Here $\mathcal{O}^{\prime }$ denotes the causal complement, $H_{R}^{\bot }$
the real orthogonal complement in the sense of the inner product $Re\left(
\psi ,\varphi \right) $ and $H_{R}^{\prime }$ is the symplectic complement
in the sense of $\func{Im}\left( \psi ,\varphi \right) .$ Representing $%
\mathcal{O}^{\prime }$ as a union of double cones, one defines $H_{R}(%
\mathcal{O}^{\prime })$ by additivity.

The direct construction of the interaction-free algebraic bosonic net for $%
(m,s=integer)$ is now achieved by converting the ''premodular'' theory of
real subspaces of the Wigner space into the Tomita-Takesaki modular theory
for nets of von Neumann algebras using the Weyl functor \cite{Ley}:

The application of the Weyl functor $\mathcal{\Gamma }$ to the net of real
spaces: 
\begin{equation}
H_{R}(\mathcal{O})\stackrel{\mathcal{\Gamma }}{\rightarrow }\mathcal{A}(%
\mathcal{O})\equiv alg\left\{ W(f)\left| f\in H_{R}(\mathcal{O})\right.
\right\}
\end{equation}
leads to a net of von Neumann algebras in $\mathcal{H}_{Fock}\,$which are in
``standard position'' with respect to the vacuum state. The ensuing a
modular theory restricted to the Fock vacuum $\Omega $ is geometric: 
\begin{eqnarray}
\Gamma (s) &=&S,\,\,\,SA\Omega =A^{*}\Omega ,\,\,\,A\in \mathcal{A}(W) \\
S &=&J\Delta ^{\frac{1}{2}},\,\,\,J=\Gamma (j),\,\,\,\Delta ^{i\tau }=\Gamma
(\delta ^{i\tau })  \nonumber
\end{eqnarray}
The proof of this theorem uses the functorial formalism of \cite{Ley}. It
should be evident from the derivation that the wedge localization concept in
Fock obtained in this functorial way from the Wigner theory only holds for
interaction free situations. The Fock space is also important for
interacting QFT, but in that case the wedge localization enters via
scattering theory as in section 4, and not just through Wigner's
representation theory.

Clearly the $W$- or $\mathcal{O}$- indexing of the Hilbert spaces
corresponds to a localization concept via modular theory. Specifically $%
H_{R}(\mathcal{O})+iH_{R}(\mathcal{O})$ is a certain closure (in the graph
topology of the operator $\Delta ^{\frac{1}{2}})$ of the one particle
component of the Reeh-Schlieder domain belonging to the localization region $%
\mathcal{O}.$ Although for general localization region the modular operators
are not geometric, there is one remaining geometric statement which presents
itself in the form of an algebraic duality property \cite{EO}: 
\begin{equation}
\mathcal{A}(\mathcal{O}^{\prime })=\mathcal{A}(\mathcal{O})^{\prime
},\,\,\,\,Haag\,\,Duality
\end{equation}
Here the prime on the von Neumann algebra has the standard meaning of
commutant. In the following we make some schematic additions and completions
which highlight the modular localization concept for more general
interaction-free theories different from the massive bosonic case.

\begin{itemize}
\item  In the case of $m\neq 0,s=$ halfinteger, the Wigner theory produces a
mismatch between the ``quantum'' ( in the sense of the commutant) and the
``geometric'' opposite of $H_{R}(W),$ which however is easily taken care of
by an additional factor $i$ (interchange of symplectic complement with real
orthogonal complement). This (via the physical localization property)
requires the application of the CAR-functor instead of the CCR-functor, as
well as the introduction of the well-known Klein transformation $K$ which
corrects the above mismatch in Fock space: 
\begin{eqnarray}
J &=&K\mathcal{F}_{CAR}(ij)K^{-1} \\
\mathcal{A}(\mathcal{O}^{\prime }) &=&K\mathcal{A}(\mathcal{O})^{\prime
}K^{-1}  \nonumber
\end{eqnarray}
where the $K$ is the twist operator of the ``twisted''\thinspace \thinspace
Haag\thinspace \thinspace \thinspace Duality \cite{Haag} and $j$ is related
to the TCP-operator as before.

\item  For $m=0,h=$(half)integer, as a consequence of the nonfaithful
representation of the zero mass little group $E(2)$ (the two-dimensional
euclidean group or rather its two-fold covering), the set of possible $u-v$
intertwiners is limited by the selection rule: $\left| A-B\right| =\pm h.$
This means on the one hand that there are e.g. no covariant intertwiners
which lead to $D^{\left[ \frac{1}{2},\frac{1}{2}\right] }$ (vector-potential
of classical Maxwell theory), $D^{\left[ \frac{1}{2},\frac{1}{2}\right]
}\otimes (D^{\left[ \frac{1}{2},0\right] }+D^{\left[ 0,\frac{1}{2}\right] })$
(Rarita-Schwinger potential for massless particles), gravitational
potentials etc. On the other hand, all local bilinear expressions in the
allowed covariant intertwiners which could serve as an inner product vanish
and hence cannot be used in order to rewrite the Wigner inner product (for
e.g. $h=1$ any local inner product in terms of field strength intertwiners $%
F_{\mu \nu }(p)$ vanishes as a consequence of the mass shell condition $%
p^{2}=0).$ A reasonable compromise consists in relaxing on strict
L-covariance and compact (double cone) modular localization, but retaining
the relation with the Wigner inner product. One then may describe the Wigner
space in terms of a vectorpotential which depends in addition on a spacelike
direction $e$ and which has the following affine Lorentz transformation: 
\begin{eqnarray}
\left( U(\Lambda )A\right) _{\mu }(p,e) &=&\Lambda _{\mu }^{\nu }A_{\nu
}(\Lambda ^{-1}p,\Lambda ^{-1}e) \\
&=&\Lambda _{\mu }^{\nu }A_{\nu }(\Lambda ^{-1}p,e)+p_{\mu }G(p,\Lambda ,e) 
\nonumber
\end{eqnarray}
where the ''gauge'' contribution $G$ by which one has to re-gauge (in order
to refer to the original spacelike polarization vector $e)$ is a nonlocal
term which follows from the above definitions. Using the x-t-boost for the
definition of $\delta ^{it}$ and defining $j$ with the help of the TCP
operation as before, one again obtains a wedge localized real subspace $%
H_{R}(W_{st})$ which contains vectorpotentials with $e$ pointing into the
wedge. This space is the same as if we would have constructed the real
modular subspace of the Wigner wave function space of right and left hand
polarized photons without vectorpotentials. After applying the Weyl functor,
we again obtain a covariant net of wedge algebras which is described in
terms of slightly nonlocal semiinfinite stringlike vectorpotential field
coordinates whose relation to the local $F_{\mu \nu }(x)$ field strength can
be shown to be given by: 
\begin{equation}
A_{\mu }(x,e)=\int_{0}^{\infty }e^{\nu }F_{\mu \nu }(x-es)ds
\end{equation}
If we now define the modular localization subspaces as before by starting
from the wedge region, we find that the (smoothened versions of) the
vectorpotentials are members of these subspaces (or their translates) as
long as the spacelike directions $e$ point inside the wedges. They are lost
if we form the localization spaces belonging to e.g. double cone regions.
Hence these stringlike localized vector potentials appear in a natural way
in our modular localization approach for the wedge regions. Whereas the
natural use of such nonpointlike objects in a future interacting theory
based on modular localization may be possible, the present formulation of
gauge theories within any known perturbative scheme is based on pointlike
fields in an Hilbert space involving ghosts and nonunitary (pseudo-unitary)
Lorentz-transformations instead of the previous affine action in Wigner
space. In this way one formally keeps the classical point localization
through the process of deformation by interaction and the ghost removal is
only done at the end of the calculation, i.e. the ghosts act like a
catalyser for forcing theories involving spin$\geq 1$ vector mesons into the
standard renormalizable framework. Such catalyzers which leave no intrinsic
observable mark in the physical results may be OK. in chemistry, but they go
a bit against the Bohr-Heisenberg spirit of removing all nonobservable
aspects from the formalism. algebraic QFT therefore considers these BRS
cohomological techniques as successful but preliminary.
\end{itemize}

\section{Thermal Aspects of Modular Localization}

In modular theory the dense set of vectors which are obtained by applying
(local) von Neumann algebras in standard position to the standard (vacuum)
vector forms a core for the Tomita operator $S.$ The domain of $S$ can then
be described in terms of the +1 (or -1) closed real subspace of $S.$ In
terms of the ``premodular'' objects $s$ in Wigner space and the modular
Tomita operators $S$ in Fock space we introduce the following nets of
wedge-localized dense subspaces: 
\begin{equation}
H_{R}(W)+iH_{R}(W)=dom(s)\subset H_{Wigner}
\end{equation}

\begin{equation}
\mathcal{H}_{R}(W)+i\mathcal{H}_{R}(W)=dom(S)\subset \mathcal{H}_{Fock}
\end{equation}

These dense subspaces become Hilbert spaces in their own right if we use the
graph norm of the Tomita operators. For the $s$-operators in Wigner space we
have: 
\begin{eqnarray}
\left( f,g\right) _{Wigner} &\rightarrow &\left( f,g\right) _{G}\equiv
\left( f,g\right) _{Wig}+\overline{\left( sf,sg\right) }_{Wig}  \label{graph}
\\
&=&\left( f,g\right) _{Wig}+\left( f,\delta g\right) _{Wig}  \nonumber
\end{eqnarray}
The graph topology insures that the wave functions are strip-analytic in the
wedge rapidity $\theta $: 
\begin{eqnarray}
p_{0} &=&m(p_{\perp })\cosh \theta ,\,\,\,\,p_{1}=m(p_{\perp })\sinh \theta
,\,\,\,m(p_{\perp })=\sqrt{m^{2}+p_{\perp }^{2}}  \label{rap} \\
&&strip:0<Imz<\pi ,\,\,\,\,\,\,z=\theta _{1}+i\theta _{2}  \nonumber
\end{eqnarray}

where this ''G-finiteness'' (\ref{graph})is precisely the analyticity
prerequisite for the validity of the KMS property for the two-point
function. For scalar Bosons we have for the Wigner inner product restricted
to the wedge : 
\begin{eqnarray}
\left( f,g\right) _{Wig}^{W} &=&\left\langle A(\overline{\hat{f}})A^{*}(\hat{%
g})\right\rangle _{0}\stackrel{KMS}{=}\left\langle A^{*}(\hat{g})\Delta A(%
\overline{\hat{f}})\right\rangle _{0} \\
&&\stackrel{CCR}{=}\left[ A^{*}(\hat{g})A(\delta \overline{\hat{f}})\right]
+(f,\delta g)_{Wig}^{W},\,\,\,\,\delta =e^{2\pi K}  \label{thermal}
\end{eqnarray}

\begin{equation}
\curvearrowright \left( f,g\right) _{Wig}^{W}\equiv \left( f,g\right)
_{K,T=2\pi }=\left[ A^{*}(\hat{g})A(\frac{\delta }{1-\delta }\overline{\hat{f%
}})\right]
\end{equation}
Here we used a field theoretic notation ($A^{*}(\hat{g})$ is a smeared
scalar complex field of the type (\ref{field}) linear in $\hat{g}$ with $%
supp.\hat{g}\in W$ ) in order to emphasize the typical thermal denominator
in the c-number commutator on the right hand side and not only implicit as
the required restriction of the wave functions to the wedge region on the
left hand side. Of course the c-number commutator may be rewritten in terms
of p-space Wigner wave functions for particles and ($\delta ^{\frac{1}{2}}$%
-transformed) antiparticles in such a way that the localization restriction
is guarantied by the property that the resulting expression is finite if the
wave functions are finite in the sense of the graph norm. With the
localization temperature in this way having been made manifest, the only
difference between localization temperatures and heat bath temperatures (for
a system enclosed in a box) on the level of field algebras in Fock space
corresponds to the difference between hyperfinite type $III_{1}$ and type $I$
von Neumann algebras. On the level of the generators of the modular group
this should correspond to a difference in their spectra. The fact that the
boost $K\,$appears instead of the Hamiltonian $H$ reveals one significant
difference between the two situations. For the heat bath temperature of a
Hamiltonian dynamics the modular operator $\delta =e^{-2\beta \mathbf{H}}$
is bounded on one particle wave functions whereas the unboundedness of $%
\delta =e^{2\pi K}$ in (\ref{thermal}) enforces the localization (strip
analyticity) of the Wigner wave functions i.e. the boost does not permit a
KMS state on the full algebra.

This difference results from the two-sided spectrum of $K$ as compared to
the boundedness from below of $H.\,$In fact localization temperatures are
inexorably linked with unbounded symmetry operators.

The generalization to fermions as well as to particles of arbitrary spin is
easily carried out. The differences between $K$ and $H$ also leads to
somewhat different energy distribution functions for small energies so that
Boson $K$-energy distributions may appear as those of $H$ heat bath Fermions
. In this context one is advised to discuss matters of statistics not in
Fourier space, but rather in spacetime where they have their unequivocal
physical interpretation.

One may of course consider KMS state on the same $C^{*}$-algebra with a
different Hawking $K$-temperature than $2\pi ;$ however, such a situation
cannot be obtained by a localizing restriction. Mathematically $C^{*}$%
-algebras to different $K$-temperatures are known to belong to different
folia (in this case after von Neumann closure to unitarily inequivalent $%
III_{1}$-algebras) of the same $C^{*}$-algebra. Or equivalently, a scaled
modular operator $\Delta ^{\alpha i\tau }$ cannot be the modular operator of
the same theory at a different temperature as it would be the case for type
I algebras. Localization temperatures are not freely variable.

For those readers who are familiar with Unruh's work we mention that the
Unruh Hamiltonian is different from $K$ by a factor $\frac{1}{a}$ where $a$
is the acceleration.

More generally we may now consider matrix elements of wedge-localized
operators between wedge localized multiparticle states. Then the KMS
property allows to move the wedge localized particle state as an
antiparticle at the analytically continued rapidity $\theta +i\pi $ from the
ket to the bra. The simplest illustration is the two-particle matrix element
of a free current of a charged scalar field $j_{\mu }(x)=:\phi ^{*}\stackrel{%
\leftrightarrow }{\partial }_{\mu }\phi :$ smeared with the wedge supported
function $\hat{h}$ (but any other free field composite would also serve)$:$%
\begin{eqnarray}
&&\left\langle 0\left| \int j_{\mu }(x)\hat{h}(x)d^{4}x\right| f,\delta ^{%
\frac{1}{2}}g^{c}\right\rangle \stackrel{KMS}{=}\left\langle 0\left| \left(
\Delta \phi ^{*}(\delta ^{\frac{1}{2}}\overline{\hat{g}})\Delta ^{-1}\right)
^{*}\int j_{\mu }(x)\hat{h}(x)d^{4}x\right| f\right\rangle   \nonumber \\
&=&\left\langle \bar{g}\left| \int j_{\mu }(x)\hat{h}(x)d^{4}x\right|
f\right\rangle   \label{cro}
\end{eqnarray}
Here $\delta ^{\frac{1}{2}}g^{c}$ is the charged transformed antiparticle in
the Wigner wave function $g$ at the analytically continued rapidity $\theta
+i\pi ,$ whereas $\hat{g}$ denotes as before the wedge-localized spacetime
smearing function whose mass shell restricted Fourier transform corresponds
to the boundary value of the analytically continuable Wigner wave function $g
$. Moving the left hand operator to the left vacuum changes the antiparticle
charge to the particle charge. Since the $H_{R}(W)+iH_{R}(W)$ complex
localization spaces are dense in the Wigner space, the momentum space kernel
for both sides of (\ref{cro}) takes the familiar form 
\begin{equation}
\left\langle p^{\prime }\left| j_{\mu }(0)\right| p\right\rangle =%
\stackunder{z\rightarrow \theta +i\pi }{anal.cont.}\left\langle 0\left|
j_{\mu }(0)\right| p,p^{^{\prime }}(z)\right\rangle   \label{cross}
\end{equation}
where $p^{\prime }(z)$ is the rapidity parametrization of above (\ref{rap}).
This famous crossing symmetry, which is known to hold also in each
perturbative order of renormalizable interacting theories, has never been
derived in sufficient generality within a nonperturbative framework of QFT.
It is to be thought of as a kind of on shell momentum space substitute for
Einstein causality and locality (and its strengthened form, called Haag
duality). As such it played an important role in finding a candidate for a
nonperturbative S-matrix of the famous Veneziano dual model. Although it
stood in this indirect way on the cradle of string theory, the recent string
theoretic inventions seem to pay little attention to these physical origins.

If crossing symmetry is really a general property of local QFT, a conjecture
(only proven in perturbation theory, as mentioned before) which nobody seems
to doubt, then it should be the on shell manifestation of the off shell KMS
property (originating from causality via the Reeh-Schlieder property) for
modular wedge localization. In the construction of wedge localized thermal
KMS states on the algebra of mass shell operators satisfying the
Zamolodchikov-Faddeev algebraic relations\footnote{%
As will become clear in the next section, although these operators are
nonlocal, they generate the wedge localized states and as a consequence the
modular KMS formalism is applicable to them.} in the momentum space rapidity 
\cite{Zam}, the derivation of crossing symmetry is similar (albeit more
involved) to the previous free field derivation \cite{S1} and the argument
can be found in section 4 of this work. Recently more general arguments
based on the Haag-Ruelle scattering theory which also hold for the case of
nonfactorizing QFT's in d=1+1 and higher dimensions were proposed \cite
{Nieder}. In our approach it turns out that the general crossing symmetry in
any dimension is indeed related to the wedge KMS condition, but that this
statement cannot be derived just from scattering theory alone. It rather
follows from the existence and the modular intertwining property of the
modular M\o ller operator $U$ (next section) which, unlike the S-matrix, is
not just an object of scattering theory as in nonrelativistic physics, but
is defined in terms of modular wedge localization. For its existence we have
to make an assumption which we presently are not able to derive within the
framework of algebraic QFT. We intend to use this object in order to prove 
\cite{SWi} the uniqueness of \textit{the main inverse problem} of QFT: $%
S_{scat}\rightarrow QFT.$

In fact the very special free field formalism of the first two sections may
be generalized into two directions:

\begin{itemize}
\item  interacting fields

\item  curved spacetime
\end{itemize}

As mentioned before, low-dimensional interacting theories will be discussed
in the next section. For the generalization to curved space time (e.g. the
Schwarzschild black hole solution) it turns out that only the existence of a
bifurcated horizon together with a certain behavior near that horizon
(``surface gravitation'') \cite{Sewell}\cite{SV} is already sufficient in
order to obtain the thermal Hawking-Unruh aspect. In the standard treatment
one needs isometries in spacetime i.e. classical horizons defined in terms
of Killing vectors. The idea of modular localization suggests to consider
also e.g. double cones for which there is no spacetime isometry but only an
isometry in $H_{Wigner}$ or $\mathcal{H}_{Fock}.$ Of course such
enlargements of spaces for obtaining a better formulation or a
generalization of a problem are commonplace in modern mathematics,
particularly in noncommutative geometry. The idea is that one replaces the
ill-defined isometries by a geometrically ``fuzzy'' but well-defined
symmetry transformations in quantum space, which only near the horizon
looses its spacetime fuzziness. The candidates for these nongeometric
symmetries are the modular automorphisms of von Neumann algebras of
arbitrary space time regions together with suitable faithful states from the
local folium of admissable states. Although the restriction of the global
vacuum state is in that folium, it is not always the appropriate state for
the construction of the modular automorphism.

In this context one obtains a good illustration by the (nongeometric)
modular theory of, e.g., the double cone algebra of a massive free field.
From the folium of states one may want to select that vector, with respect
to which the algebra has a least fuzzy (most geometric) behavior under the
action of the modular group. Appealing to the net subtended by spheres $S$
at time t=0 one realizes that algebras localized in these spheres are
independent of the mass. Since m=0 leads to a geometric modular situation%
\footnote{%
The modular group is a one-parametric subgroup of the conformal group.} for
the pair ($\mathcal{A}_{m=0}(S),\Omega _{m=0}),$ and since the nonlocality
of the modular group of the massive theory in the ``wrong'' (massless)
vacuum ($\mathcal{A}_{m\neq 0}(C(S)),$ $\Omega _{m=0})$ in the subtended
double cone $C(S)$ is only the result of the fuzzy propagation inside the
light cone (the breakdown of Huygens principle or the ``reverberation''
phenomenon caused by the mass), the fuzziness of the modular group for this
pair is a pure propagation phenomenon, i.e., can be understood in terms of
the deviation from Huygens principle. In view of the recent micro-local
spectrum condition, one expects this nonlocal modular group action even in
the correct vacuum i.e. for $(A_{m\neq 0}(C(S)),\Omega _{m\neq 0})$ to have
modular groups\footnote{%
They differ from the previous situation by a Connes cocycle \cite{Haag}.}
whose generators are pseudo-differential instead of (local) differential
operators \cite{Fred}. In this case the asymptotically local action near the
light-like horizon will continue to hold. In order to avoid the pathology of
the d=1+1 scalar zero mass field, one should use for the above consideration
a massive free spinor field whose massless limit gives a two-component field
with the first component only depending on the left light cone and the
second on the right hand light cone. In fact the above observation is very
much related to the zero mass theory results from Sewell's restriction \cite
{Sewell} to the light cone horizon (boundary). It is my conviction that all
the recent speculations about the quantum version of Bekenstein's classical
entropy and in particular the horizon (``holographic'') aspects of the
associated degrees of freedom are manifestations of modular properties of
generic nonperturbative QFT which however are overlooked in the Lagrangian
quantization method. This and similar subjects will be the content of a
separate paper with Wiesbrock.\cite{SWi}.

The Hilbert space setting of modular localization offers also a deeper
physical understanding of the universal domain $\mathcal{D}$ which plays a
rather technical role in the Wightman framework \cite{Wig Str} In the
modular localization approach the necessity for such a domain appears if one
wants to come from the net of localization spaces which receive their
natural topology from the (graphs) net of Tomita operators $\bar{S}(\mathcal{%
O})$ to a net of (unbounded) polynomial algebras $\mathcal{P}(\mathcal{O})$
such that: 
\begin{equation}
dom\,\,\bar{S}(\mathcal{O})\cap \mathcal{D}=\mathcal{P}(\mathcal{O})\Omega
=dom\,\,\mathcal{P}(\mathcal{O});
\end{equation}
This domain is of course also expected to be equal to $\mathcal{A}(\mathcal{O%
})\Omega .$ Here we used a more precise notation which distinguishes between
the operator $S$ defined on the core $\mathcal{A}(0)\Omega $ and its closure 
$\bar{S}$ which is defined on $\mathcal{H}_{R}(\mathcal{O})+i\mathcal{H}_{R}(%
\mathcal{O}).$

\section{Modular Wedge-Localization and Factorizing Theories}

In this section we will show that the modular localization can be used as a
starting point for constructing for interacting theories. This constructive
approach is presently most clear in the case of factorizing d=1+1 QFT's to
which we will limit ourselves in this section. We remind the reader that
``factorizable'' in the intrinsic physical interpretation \cite{schroer}\cite
{S2} of algebraic QFT means that the long distance limit of the S-matrix
(which only consists of the two-particle elastic part and automatically 
fulfills the Yang-Baxter relation as a physical consistency condition)
defines a QFT model in its own right which we call ``factorizable''. It has
no on-shell particle creation, but it does possess a rich off-shell (or
``virtual'') particle structure i.e. it is a full-fledged QFT with
nontrivial vacuum polarization caused by interactions. It should be viewed
as being the simplest (no real creation) representative of a vast
equivalence class of complicated (with real particle creation) models which
share the same particle content together with the same long distance
S-Matrix $S_{lim}.$ This notion of factorizable is better suited for the
present use in local quantum physics than the traditional ``integrability''
which is defined via quantization.. The idea is somewhat analogous to the
construction of  the simplest representative in a  long-distance equivalence
class (in the sense of the S-matrix) of a given superselection class. In
other words each general d=1+1 field theory has an asymptotic companion
which has the same superselection sectors ($\simeq $ same particle structure
or incoming Fock space), but vastly simplified dynamics associated to a
factorizing S-matrix . In d=3+1 this distinguished representative reduces to
a free field with the same superselection structure as the other members of
the equivalence class. This intrinsic understanding without imposing
conservation laws or even Yang-Baxter structures (but rather obtaining them
from consistency of the long distance limits of scattering operators) gives
an enhanced significance of factorizable models as the simplest
representative in a class of general models with the same charge
superselection structure. This is of course analogous to the physical
significance of the short distance universality class of conformal field
theories. The present field theoretic understanding of the
bootstrap-formfactor program, which is largely a collection of more or less
plausible (but very successful) recipes \cite{Smi}, leaves a certain amount
of conceptual clarity to be desired. Our main concern in this paper is
therefore the elaboration of a framework which positions the
formfactor-bootstrap program within general QFT in such a way that it can be
used  as a theoretical laboratory for the latter.

All applications of modular localization to interacting theories are based
on the observation that in asymptotically complete theories with a mass gap,
the full interaction resides in the Tomita operator $J(W),$ whereas the
modular group $\Delta ^{i\tau }(W)$ for wedges (being equal to Lorentz
boosts) as well as the continuous Poincar\'{e} group transformations is
``blind'' against interactions (the physical representation of the
Poincar\'{e} group are already correctly defined on the free incoming
states). In fact the interaction resides in those disconnected parts of the
Poincar\'{e} group which involve antiunitary time reflections and the Tomita 
$J$ for the standard wedge (containing the origin in its edge) inherits this
from the TCP operator. To be more explicit the Haag Ruelle scattering theory
together with the asymptotic completeness easily yield (for each wedge): 
\begin{equation}
J=S_{s}J_{0},\,\,\,\,\Delta ^{i\tau }=\Delta _{0}^{i\tau }
\end{equation}
where the subscript $0$ refers to the free incoming situation and we have
omitted the reference to the particular wedge. It is very important here to
emphasize that this interpretation of modular data in terms of scattering
theories exists \textit{only for wedge regions}. It enters via the TCP
invariance and is not just a consequence of Haag-Ruelle scattering theory,
since the latter cannot be formulated within wedges. At this point we differ
from the approach in \cite{Nieder} which seeks to extract the cyclic
relation associated with the crossing symmetry from the KMS condition of the
wedge algebra of the \textit{interacting local fields}, whereas we will
generate this algebra from semilocal operators without any additional large
time limits. In fact the cyclic equation is equivalent to the KMS relation
for the interacting wedge algebra written in special nonlocal generators.
These nonlocal operators are on-shell\footnote{%
For this reason they cannot be used to construct better than wedge
localizations. Compactly localized algebras, as double cone algebras, have
to be constructed with the help of intersections and have new off-shell
generators.} and free of vacuum polarizations. Their rapidity space
creation- and annihilation- components are forming the Zamolodchikov-Faddeev
algebra. We will call them (vacuum-polarization-) free wedge generators,
abbreviated FWG, because they are on-shell operators. They are in some sense
a nonlocal generalization of free fields and agree with the latter in case
of absence of interactions, but they do generate the interacting wedge
algebra and therefore merit the attribute ``semilocal'', i.e. the FWG`s are
special semilocal operators. These objects had been introduced already in
previous publications of the author \cite{S2}\cite{schroer}; here we will
present them in more details and also use the opportunity to correct some
earlier errors.

No physical interpretation is yet known for the modular objects of compactly
localized algebras of e.g. double cones. One expects in that case, that
different from the wedge case, not only the reflection $J,$ but also the
modular group $\Delta ^{i\tau }$ will depend on the interaction. As will be
seen later, the double cone algebras can be constructed from the wedge
algebras.

In order not to change the traditional notation $S$ for the Tomita
involutions, we use the subscript $s$ whenever we mean the S-matrix of
scattering theory. The most convenient form for the previous equation which
puts the modular aspect of scattering in evidence is: 
\begin{equation}
S=S_{s}S_{0}
\end{equation}
where $S$ and $S_{0}$ are the antiunitary Tomita operators and $S_{s}$ is
the unitary scattering operator. Therefore the scattering operator in
relativistic QFT has two rather independent aspects: it is a global operator
in the sense of large time limits of scattering theory, and it has a modular
localization interpretation in measuring the deviation of $J$ or $S$ from
their free field values $J_{0},S_{0}$ i.e. \textit{it is a relative modular
invariant}. This modular aspect is characteristic of local quantum physics
and has no counterpart in nonrelativistic theory or quantum mechanics.

The modular subspace of $\mathcal{H}_{Fock}\equiv \mathcal{H}_{in\text{ }}$%
for the standard wedge is characterized in terms of the following equations%
\footnote{%
The distiction between the $\pm $sign is not very important since the
multiplication with $i$ converts one real subspace into the other.}: 
\begin{eqnarray}
S_{s}S_{0}\mathcal{H}_{R} &=&\mathcal{H}_{R} \\
S_{s}S_{0}\psi &=&\psi ,\,\,\,\psi \in \mathcal{H}_{R}  \nonumber
\end{eqnarray}
There is one more important idea which is borrowed from scattering theory
namely the existence of a ``modular M\o ller operator'' \cite{S2} $U$ which
is related to the S-matrix as: 
\begin{equation}
S_{s}=UJ_{0}U^{*}J_{0}
\end{equation}
This correspond to the well-known standard formula $S_{s}=(\Omega
^{out})^{*}\Omega ^{in}.$ Writing the S-matrix in terms of an Hermitian
phase matrix $\eta $ as $S=e^{i\eta }$ it is not difficult to find an $U$ in
terms of $\eta .$ Note however that the Haag-Ruelle scattering theory (as
well as its more formal but better known LSZ predecessor) in local quantum
physics does not provide a M\o ller isometry between Heisenberg states and
incoming states because the scattering state space and the space for the
interacting fields are identical since in local quantum physics, different
from the nonrelativistic rearrangement scattering theory, one does not
introduce a separate space of asymptotic fragments.

The idea of introducing such an object into our modular approach comes from
the unitary equivalence of the interacting and the free hyperfinite type $%
III_{1}$ wedge algebras: 
\begin{equation}
\mathcal{A}(W)=U\mathcal{A}^{in}(W)U^{*}  \label{Mo}
\end{equation}
We demand the $U$-invariance of the vacuum $U\Omega =\Omega $ and the
one-particles space $U\mathcal{H}^{(1)}=\mathcal{H}^{(1)}.$ The physical
idea behind this is that whereas the vacuum and the one-particle states
cannot be resolved from the rest of the energy-momentum spectrum in compact
regions as e. g. double cones, the semiinfinite wedge region, which is left
invariant by the associated Lorentz-boosts, does allow such a resolution.
Note that this situation is different from the problem of unitary
equivalence of the canonical equal time commutation relations in the free
versus the interacting case. A unitary equivalence in this case (of an
algebra belonging to a region with trivial spacelike complement) would be
forbidden by Haag's theorem \cite{Haag} on the nonexistence of the
interaction picture in QFT.

The above characterization of $U$ may be replaced by a slightly more
convenient one\footnote{%
I am indebted to H.-W. Wiesbrock for emphasizing this intertwining property
as the most convenient definition of $U.$} in terms of an intertwining
property between modular operators: 
\begin{equation}
US_{0}=SU  \label{U}
\end{equation}
In terms of localized spaces, the $U$ has the property: 
\begin{equation}
U\mathcal{H}_{R}^{in}(W)=\mathcal{H}_{R}(W)  \label{sp}
\end{equation}
Unfortunately one cannot conclude from this transformation of spaces (\ref
{sp}) (which is the only property derivable from (\ref{U})) in favor of the
validity of (\ref{Mo}) and therefore the spatial modular property is too
weak for the construction of the wedge algebra. In order to find another
method for constructing $\mathcal{A}(W),$ we first study a simple model.

Between the two simplest possibilities, the Ising field theory with $%
S_{s}^{(2)}=-1$ and the (non-parity invariant) Federbush model with $%
S_{I,II}^{(2)}=e^{i\pi g}$ we chose the latter because it allows also a
Lagrangian interpretation and hence is simpler to describe to readers
familiar with the Lagrangian quantization approach to QFT. The model
consists in coupling two species of Dirac fermions via a (parity violating)
current-pseudocurrent coupling \cite{Wigh}\cite{SW}: 
\begin{equation}
\mathcal{L}_{int}=g:j_{\mu }^{I}j_{\nu }^{II}:\varepsilon ^{\mu \nu
},\,\,\,\,j_{\mu }=:\bar{\psi}\gamma _{\mu }\psi :
\end{equation}
One easily verifies that: 
\begin{eqnarray}
\psi _{I}(x) &=&\psi _{I}^{(0)}(x)\vdots e^{ig\Phi _{II}^{(l)}(x)}\vdots
\label{local} \\
\psi _{II}(x) &=&\psi _{II}^{(0)}(x)\vdots e^{ig\Phi _{I}^{(r)}(x)}\vdots 
\nonumber \\
\psi _{I,II}^{(0)}(x) &=&\frac{1}{\sqrt{2\pi }}\int \left(
e^{-ipx}a_{I,II}(\theta )+e^{ipx}b_{I,II}^{*}(\theta )\right) d\theta
\end{eqnarray}
where $\Phi ^{(l,r)}=\int_{x^{\prime }\lessgtr x}j_{0}dx^{\prime }$ is a
potential of $j_{\mu 5}$ i.e. $\partial _{\mu }\Phi \sim \varepsilon _{\mu
\nu }j^{\nu }=j_{\mu 5}$ and the superscript $l,r$ refers to whether we
choose the integration region for the line integral on the spacelike left or
right of $x$. The triple ordering is needed in order to keep the closest
possible connection with classical geometry and localization and in
particular to maintain the validity of the classical field equation in the
quantum theory; for its meaning and its conversion into the standard Fermion
Wick-ordering we refer to the above papers. This conceptually simpler triple
ordering can be recast into the form of the analytically (computational)
simpler standard Fermion Wick-ordering in terms of the (anti)particle
creation/annihilation operators $a_{I,II}^{\#}(\theta ),b_{I,II}^{\#}(\theta
)$. Although in this latter description the classical appearance of locality
is lost, the quantum exponential do still define local Fermi-fields\cite{SW}%
; in the case of relative commutation of $\psi _{I}$ with $\psi _{II}$ the
contributions from the exponential (disorder fields) compensate. This model
belongs to the simplest class of factorizing models (those with rapidity
independent S-matrix) and its explicit construction via the formfactor
program is almost identical to that of the massive Ising field theory \cite
{Karowski}. The reason why it does not appear under this approach in the
literature is that the bootstrap classification was limited to strictly
parity conserving theories. For our present purposes it serves as the
simplest nontrivial illustration of new concepts arising from modular
localization.

Despite the involved looking local fields (\ref{local}), the wedge algebras
are easily shown to be of utmost simplicity: 
\begin{eqnarray}
\mathcal{A}(W) &=&alg\left\{ \psi _{I}^{(0)}(f)U_{II}(g),\psi
_{II}^{(0)}(h);suppf,h\in W\right\}  \label{alg} \\
\mathcal{A}(W^{\prime }) &=&\mathcal{A}(W)_{Klein}^{^{\prime }}=alg\left\{
\psi _{I}^{(0)}(f),\psi _{II}^{(0)}(h)U_{I}(g);suppf,h\in W^{\prime }\right\}
\nonumber
\end{eqnarray}
i.e. the two wedge-localized algebras ($W$ denotes the right wedge) are
generated by free fields ``twisted'' by global $U(1)$-symmetry
transformation of angle $g$ (coupling constant) and the subscript ``Klein''
denotes the well-known Klein transformation associated with the $2\pi $
Fermion rotation. The right hand side follows from the observation that with 
$x$ restricted to $W$, one may replace the exponential in $\psi _{I}$ in (%
\ref{local}) (which represents a left half space rotation) by the full
rotation since the exponential of the right half-space charge is already
contained in the right free fermion algebra etc. The following unitarily
equivalent description of the pair $A(W),A(W^{\prime })$ has a more
symmetric appearance under the extended parity symmetry $\psi
_{I}(t,x)\leftrightarrow \psi _{II}(t,-x)$:\vspace{0in} 
\begin{eqnarray}
\mathcal{A}(W) &=&alg\left\{ \psi _{I}^{(0)}(f)U_{II}(\frac{g}{2}),\psi
_{II}^{(0)}(h)U_{I}(-\frac{g}{2});suppf,h\in W\right\}  \label{rep} \\
\mathcal{A}(W^{\prime }) &=&alg\left\{ \psi _{I}^{(0)}(h)U_{II}(\frac{g}{2}%
),\psi _{II}^{(0)}(f)U_{I}(-\frac{g}{2});suppf,h\in W^{\prime }\right\} 
\nonumber
\end{eqnarray}
The verification of all these properties is elementary, and no modular
theory is needed. The computation \cite{SW} of the scattering matrix $S_{s}$
from (\ref{local}) is most conveniently done by Haag-Ruelle scattering
theory \cite{Haag}: 
\begin{eqnarray}
S_{s}\left| \theta _{1}^{I},\theta _{2}^{II}\right\rangle
&=&S_{s}^{(2)}\left| \theta _{1}^{I},\theta _{2}^{II}\right\rangle =e^{i\pi
g}\left| \theta _{1}^{I},\theta _{2}^{II}\right\rangle \\
S_{s}^{(n)} &=&\prod_{pairings}S_{s}^{(2)}  \nonumber
\end{eqnarray}
These formulae (including antiparticles) can be collected into an operator
expression \cite{SW} : 
\begin{equation}
S_{s}=\exp i\pi g\int \rho _{I}(\theta _{1})\rho _{II}(\theta
_{2})\varepsilon (\theta _{1}-\theta _{2})d\theta _{1}d\theta _{2}
\label{Fed}
\end{equation}
Where $\rho _{I,II}$ are the momentum space charge densities in the rapidity
parametrization.

The surprising simplicity of the wedge algebra of this model as compared to,
say, its double cone algebras consists in the fact that one can choose
on-shell generators. We will show that modular wedge localization for
factorizing models always leads to on shell generators for the localized
spaces, though for rapidity dependent S-matrices they do not generate the
wedge algebras.

It would now be easy to solve the n-particle modular localization equation%
\footnote{%
The Tomita operator $S$ for Fermions is different from that of Bosons by a
Klein transformation. For a special family of d=1+1 solitons the correct TCP
operator has been computed by Rehren \cite{Rehren}. Since all the known
families of factorizing models are described by Fermions and Bosons and
since it is not clear whether this generalization is compatible with the
factorization we will ignore this more general TCP-situation in the present
context.}: 
\begin{eqnarray}
S\mathcal{H}_{R}^{(n)}(W) &=&\mathcal{H}_{R}^{(n)}(W);\,\,\,\mathcal{H}%
_{R}(W)=\oplus _{n}\mathcal{H}_{R}^{(n)}(W)  \label{F} \\
\mathcal{H}_{R}^{(n)}(W) &=&\left\{ \int F(\theta _{1},\theta
_{2},...,\theta _{n})\left| \theta _{1},\theta _{2},...,\theta
_{n}\right\rangle d\theta _{1}d\theta _{2}...d\theta _{n}\mid F\in
H_{strip}^{(n)}\right\}  \nonumber
\end{eqnarray}
Here $H_{strip}^{(n)}$ denotes the closure of the space of square integrable
function which allow an analytic continuation into the strip $0<Imz_{i}<\pi
,i=1...n$ and fulfill certain boundary conditions.$.$ This is just the
p-space on shell analyticity which comes from the wedge localization. In
analogy to Wightman functions in x-space, the $n!$ different boundary
prescriptions $Imz_{i_{1}}>Imz_{i_{2}}>....>Imz_{i_{n}}\longrightarrow 0$
yield the generally $n!$ different boundary values $F(\theta _{i_{1}},\theta
_{i_{2}},...\theta _{i_{n}})$ which are described by one masterfunction $F.$
Similar statements hold for the boundary values on the upper rim. In the
free case i.e. for $S_{s}=1$, one just recovers the Bose/Fermi statistics
but with the Federbush S-matrix the space consists of strip-analytic
functions which are a solution of a Riemann-Hilbert boundary problem. The
general solution of this problem (i.e. the characterization of the subspace $%
\mathcal{H}_{R}(W)$ within the full multiparticle wave function space) may
be presented as a product of special solution of the Riemann-Hilbert problem
with the general solution of the interaction free problem in $\mathcal{H}%
_{R}(W)^{in}.$ A convenient way to write the solution consists in using
auxiliary operators $Z$ as generators for a basis of state vectors: 
\begin{eqnarray}
\int d^{2}x_{1}...d^{2}x_{n}\hat{f}_{n}(x_{1},....x_{n})
&:&Z_{I,II}(x_{1})....Z_{I,II}(x_{n}):\Omega ,\,\,  \nonumber \\
\,\,supp\hat{f}_{n} &\in &W^{\otimes n},\,\,\hat{f}_{n}\,\,real  \label{form}
\end{eqnarray}
where the $Z^{\prime }s$ are on shell operators whose frequency positive and
negative momentum space components have to fulfill commutation relations
which must be compatible with the boundary relations governed by products of
two particle S-matrices. One immediately realizes that this leads to the
Zamolodchikov-Faddeev algebra relations for the Federbush S-matrix: 
\begin{equation}
Z_{I,II}(x)=\frac{1}{\sqrt{2\pi }}\int \left( e^{-ipx}c_{I,II}(\theta
)+e^{ipx}d_{I,II}^{*}(\theta )\right) d\theta  \label{Z}
\end{equation}
where the $c$ and the corresponding anti $d$ can be formally expressed in
terms of the incoming (anti)particle creation and annihilation operators: 
\begin{eqnarray}
c_{I,II}(\theta ) &=&a_{I,II}(\theta ):e^{-i\pi g\int_{-\infty }^{\theta
}\rho _{II,I}(\theta ^{\prime })d\theta ^{\prime }}: \\
d_{I,II}(\theta ) &=&b_{I,II}(\theta ):e^{i\pi g\int_{-\infty }^{\theta
}\rho _{II,I}(\theta ^{\prime })d\theta ^{\prime }}:  \nonumber
\end{eqnarray}
with the Zamolodchikov-Faddeev relations between the type I and II
particles: 
\begin{eqnarray}
c_{I}(\theta _{1})c_{II}(\theta _{2}) &=&-S^{(2)}(\theta _{1}-\theta
_{2})c_{II}(\theta _{2})c_{I}(\theta _{1})  \label{c} \\
c_{I}(\theta _{1})c_{II}^{*}(\theta _{2}) &=&-S^{(2)}(\theta _{1}-\theta
_{2})^{-1}c_{II}^{*}(\theta _{2})c_{I}(\theta _{1})  \nonumber \\
d_{I}(\theta _{1})d_{II}(\theta _{2}) &=&-S^{(2)}(\theta _{1}-\theta
_{2})d_{II}(\theta _{2})d_{I}(\theta _{1})  \nonumber \\
d_{I}(\theta _{1})d_{II}^{*}(\theta _{2}) &=&-S^{(2)}(\theta _{1}-\theta
_{2})^{-1}d_{II}^{*}(\theta _{2})d_{I}(\theta _{1})  \nonumber \\
d_{I,II}(\theta _{1})c_{II,I}(\theta _{2}) &=&-S^{(2)}(\theta _{1}-\theta
_{2})^{-1}c_{II,I}(\theta _{2})d_{I,II}(\theta _{1})  \nonumber \\
d_{I,II}(\theta _{1})c_{II,I}^{*}(\theta _{2}) &=&-S^{(2)}(\theta
_{1}-\theta _{2})c_{II,I}^{*}(\theta _{2})d_{I,II}(\theta _{1})  \nonumber
\end{eqnarray}
and the free Fermion anticommutation relations between the same type. The
simplicity of the model is reflected in the fact that interactions only take
place between species I and II and the independence of $S^{(2)}$ on $\theta
. $ It is now easy to check that the solution of the wedge localization
equation (\ref{F}) has indeed the form (\ref{form}). The use of the Z-basis
takes care of the boundary conditions in an analogous way as the
(anti)symmetry of the coefficient functions follows from the commutation
relations of the Fermion/Boson operator basis. In particular it takes care
of the inversion of the order in Z-products under Hermitian conjugation: $%
SA\Omega =A^{*}\Omega ,$ with $A$ an operator as in (\ref{form}). as well as
the coalescence of the x-space commutation relations for (\ref{Z}) with
those of (\ref{alg}) or (\ref{rep}). The interaction does make a distinction
between left and right and parity is only conserved if one also interchanges
the two species. The wedge localized states fulfill the following thermal
KMS condition

\begin{equation}
\varphi (t)=\left\langle \psi _{2}\mid \Delta ^{it}\psi _{1}\right\rangle
,\,\,\;\varphi (t-i)=\left\langle \Delta ^{it}\psi _{1}\mid \psi
_{2}\right\rangle
\end{equation}
where $\psi _{1,2}$ are n-particle localized states of the form (\ref{form})
and $\varphi $ is analytic in the open t-strip and continuous on the
boundary.

A calculation of the $J$-transformed operators $Z^{J}=JZJ$ with

\begin{eqnarray}
c_{I,II}^{J}(\theta ) &=&a_{I,II}(\theta ):e^{-i\pi g\int_{\theta }^{\infty
}\rho _{II,I}(\theta ^{\prime })d\theta ^{\prime }}: \\
d_{I,II}^{J}(\theta ) &=&b_{I,II}(\theta ):e^{i\pi g\int_{\theta }^{\infty
}\rho _{II,I}(\theta ^{\prime })d\theta ^{\prime }}:  \nonumber
\end{eqnarray}
reveals that $Z^{J}$ commutes with $Z$. This is a speciality of models with
rapidity independent S-matrices. In such models the wedge locality is
manifest since 
\begin{equation}
\left[ Z_{I}^{J\#}(\theta ),Z_{II}^{\#}(\theta ^{\prime })\right] =0
\end{equation}
and all other commutators between operators $Z^{J}$ and $Z$ of the same type
are like those of free fields. This means that $Z^{J}(x)$ commutes with $Z(y)
$ for $x,y\in W$ , even though the $Z^{\#}(x)^{\prime }s$ are nonlocal
fields.

For more general factorizing models however the verification of wedge
locality of the FWG operators is more subtle. We illustrate the procedure
for a factorizing model with one interacting (selfconjugate) particle. In
that case the $Z$ has the following form in terms of the scattering phase
shift $\vartheta $%
\begin{eqnarray}
Z(x) &=&\frac{1}{\sqrt{2\pi }}\int \left\{ e^{-ipx}Z(\theta )+h.c.\right\}
d\theta ,\,\,p=m(ch\theta ,sh\theta ) \\
Z(\theta ) &=&a(\theta )e^{-i\int_{-\infty }^{\theta }\vartheta (\theta
-\theta ^{\prime })a^{*}(\theta ^{\prime })a(\theta ^{\prime })d\theta
^{\prime }},\,\,\,S_{sc}(\theta )=e^{i\vartheta (\theta )}  \nonumber
\end{eqnarray}

\begin{eqnarray}
Z^{\#}(\theta _{1})Z^{\#}(\theta _{2}) &=&S_{sc}(\theta _{1}-\theta
_{2})Z^{\#}(\theta _{2})Z^{\#}(\theta _{1}),\,\,Z^{\#}\equiv Z\,\,or\,\,Z^{*}
\\
Z(\theta _{1})Z^{*}(\theta _{2}) &=&S_{sc}^{-1}(\theta _{1}-\theta
_{2})Z^{*}(\theta _{2})Z(\theta _{1})+\delta (\theta _{1}-\theta _{2}) 
\nonumber
\end{eqnarray}
For the following we find it convenient to introduce the path notation which
allows to denote rapidity space creation and annihilation operators by one
symbol. We write for free fields $A(x)$ smeared with wedge supported test
functions 
\begin{eqnarray}
A(\hat{f}) &=&\int f(\theta )a(\theta )+\int f(\theta -i\pi )a(\theta -i\pi
)\equiv \int_{C}f(\theta )a(\theta )  \label{wedge} \\
&&a(\theta -i\pi )\equiv a^{*}(\theta ),\,\,f(\theta -i\pi )=\bar{f}(\theta )
\nonumber
\end{eqnarray}
$C$ consists of the real $\theta -$axis and the parallel path shifted down
by $-i\pi $ and it is only the function $f$ which is analytic in the strip $%
-\pi <Im\theta <0$ and not the operators. The analyticity of $\hat{f}$ is
equivalent to the localization property. We will use the same notation for
the FWG $Z(\hat{f})$. In the application to the vacuum of course only the
creation contribution from the lower rim of the strip survives.

We want to prove that the on-shell $Z$ is a FWG i.e. the smeared operator $Z(%
\hat{f})=\int Z(x)\hat{f}(x)d^{2}x$ with supp$\hat{f}\in W$ generates the $%
^{*}$-algebra of the interacting theory localized in the wedge. In formula 
\begin{eqnarray}
\left[ JZ(\hat{f})J,Z(\hat{g})\right]  &=&0,\,\,\,\sup p\hat{f},\hat{g}\in W
\label{w.l.} \\
Z(\hat{f})\Omega  &=&\int d\theta \bar{f}(\theta )a^{*}(\theta )\Omega  
\nonumber
\end{eqnarray}

To prove this one first notices that the $Z(\theta )^{\#}$ commutes with the 
$JZ(\theta )^{\#}J$ underneath the Wick-ordering$.$ The reason is that the
exponential involve integrals over number density which extend over
complementary rapidity regions. The numerical phase factors which originate
from the commutation of these exponential factors with the $a^{\#\prime }s$
mutually compensate. There remains the contraction between the
pre-exponential $a^{\#\prime }s$ which leads to 
\begin{equation}
\int \bar{f}(\theta )g(\theta -i\pi )\exp i\int \delta _{sc}(\theta -\theta
^{\prime })n(\theta ^{\prime })d\theta ^{\prime }
\end{equation}
Shifting the integration by $i\pi ,$ and using the crossing symmetry in the
form: $\delta _{sc}(\xi +i\pi )=\delta _{sc}(-\xi )+2\pi ni,$ we see that
this contraction is equal to that in the opposite order (which has the
negative exponential). The pointwise equation $JA(-x_{0},x)J=PA(x)P$ for the
local field $A(x),$ with $P$ being the generator of the parity
transformation, is to be replaced by $P\left( \int Z(\theta )f(\theta
)d\theta \right) P\in J\mathcal{A}(W)J,$ a relation which is easily checked
using the explicit form of the Z-operators.

Now we come to the crucial part of the modular localization method, the
exploration of consequences of the KMS condition for the Z-correlation
functions. As a typical case we consider the 4-point function.

\begin{eqnarray}
\left( \Omega ,Z(f_{1^{^{\prime }}})Z(f_{2^{^{\prime
}}})Z(f_{2})Z(f_{1})\Omega \right)  &\equiv &\left\langle Z(f_{1^{^{\prime
}}})Z(f_{2^{^{\prime }}})Z(f_{2})Z(f_{1})\right\rangle _{therm} \\
&=&\left\langle Z(f_{2^{^{\prime }}})Z(f_{2})Z(f_{1})Z(f_{1^{^{\prime
}}}^{2\pi })\right\rangle _{therm}  \nonumber
\end{eqnarray}
Each side is the sum of two terms, the direct term associated with 
\begin{eqnarray}
Z(f_{2})Z(f_{1})\Omega  &=&\int f_{2}(\theta _{2}-i\pi )f_{1}(\theta
_{1}-i\pi )Z^{*}(\theta _{1})Z^{*}(\theta _{2})\Omega +c\,\,\,number\cdot
\Omega  \\
&=&\int f_{2}(\theta _{2}-i\pi )f_{1}(\theta _{1}-i\pi )S(\theta _{2}-\theta
_{1})a^{*}(\theta _{1})a^{*}(\theta _{2})\Omega +c\Omega   \nonumber
\end{eqnarray}
and the analogous formula for the bra-vector. For the inner product there
are two contraction terms consisting of direct and crossed contraction (in
indices 1 or 2) of the $a^{\#}s.$ Only the second one gives an S-matrix
factor in the integrand. The c-number term an the left hand side cancels the
direct term on the right hand side. The equality of the crossed terms on
both sides gives (using the denseness of the analytic wave functions) 
\begin{equation}
S(\theta _{2}-\theta _{1})=S(\theta _{1}-\theta _{1^{\prime }}+i\pi )\mid
_{\theta _{1^{\prime }}=\theta _{2}}
\end{equation}
i.e. one obtains the above crossing relation for the two particle S-matrix.
Higher inner products involve products of S-matrices, and it is easy to see
that the KMS condition for the FWG algebra is equivalent to the crossing
property of the S-matrix. The presence of additional local operators $A$
which can be a fortiori localized in the wedge does not influence the
validity of the KMS condition. 
\begin{eqnarray}
&&\left\langle Z(f_{1^{^{\prime }}})Z(f_{2^{\prime }})...Z(f_{m^{^{\prime
}}})AZ(f_{n})...Z(f_{2})Z(f_{1})\right\rangle _{therm}= \\
&=&\left\langle Z(f_{2^{\prime
}})..Z(f_{m})AZ(f_{n})..Z(f_{2})Z(f_{1})Z(f_{1^{^{\prime }}}^{2\pi
})\right\rangle _{therm}  \nonumber
\end{eqnarray}
The rapidity space formulation of this KMS condition is (again using
denseness of wave functions, the derivation is completely analogous to the
previous case) the desired cyclicity relation for the formfactor of that
local operator $A$ (its coefficient functions in the sense of (\ref{A}) below%
$:$ 
\begin{equation}
a_{n}(\theta _{1},\theta _{2},...,\theta _{n})=a_{n}(\theta _{2},...,\theta
_{n},\theta _{1}-2\pi i)  \label{cyc}
\end{equation}

This equation was hitherto derived as a special consequence of the crossing
symmetry for factorizing systems. The crossing symmetry itself in turn
follows from the LSZ scattering theory together with certain analytic
assumptions (\cite{BFK}) which are presently only controllable in the
factorizing setting. Without the mediation of the semilocal FWG $Z^{\#}$ it
would not have been possible to link the cyclicity equation with the KMS
property of $\mathcal{A}(W)$ i.e. the KMS relation only takes the form of
the cyclicity relation in the semilocal FWG field ``coordinates''. To the
extend that the derivation in (\cite{Nieder}) is correct, it must implicitly
contain the FGW operators.

Contrary to the previous free case, the sharpening of the support of the
test function does not improve the localization within the wedge. This is
equivalent to the statement that the reflection with $J$ does not create an
operator which is localized at the geometrically mirrored support region of
the test function $\hat{f}$ in the opposite wedge $W^{\prime }.$ It only
fulfills the commutation relation with respect to the full $W^{\prime }.$ In
fact the breakdown of parity covariance is important for the existence of
such nonlocal but wedge-localized fields, since fields which are covariant
under all transformations are expected to be either point local or
completely delocalized (i.e. not even in a wedge). We will later see that
any sharper localization requires the operator to be an infinite power
series in the $Z^{\prime }s:$%
\begin{equation}
A=\sum \frac{1}{n!}\int_{C}...\int_{C}a_{n}(\theta _{n},\theta
_{n-1},....\theta _{1}):Z(\theta _{1})....Z(\theta _{n}):  \label{A}
\end{equation}
where we again used the previously explained path notation. The sharper
localization leads to relations between the $a_{n}^{\prime }s$ $.$ Note that
since the commutations of $Z^{\prime }s$ produce S-matrix phase factors, the 
$a_{n}$ must compensate this phase factors upon commuting $\theta ^{\prime
}s.$ 
\begin{equation}
a_{n}(\theta _{1},...\theta _{i},\theta _{i+1},..\theta _{n})=S_{sc}(\theta
_{i}-\theta _{i+1})a_{n}(\theta _{1},...\theta _{i+1},\theta _{i},..\theta
_{n})  \label{co}
\end{equation}
In this respect the phase factors are like statistics terms. However since
the $a_{n}$ have meromorphic properties in the multi-$\theta $ strip
(actually for compact localization the meromorphy region is much bigger),
these phase factors must be consistent with the univaluedness in the
analytic domain. Together with other requirements, this leads to a
multi-variable Riemann-Hilbert problem for the $a_{n}^{\prime }s$ (\cite{BFK}%
). Since the d=1+1 double cone algebras $\mathcal{A}(\mathcal{O})$ are
obtained by translations and intersections 
\begin{equation}
\mathcal{A}(\mathcal{O}_{a})=U(-\frac{a}{2})\mathcal{A}(W)U(\frac{a}{2})\cap
U(\frac{a}{2})\mathcal{A}(W^{\prime })U(-\frac{a}{2})
\end{equation}
this sharper localization leads to additional restrictions for the $%
a_{n}^{\prime }s.$ For a detailed characterization of these algebras in
terms of generators I refer to forthcoming joint work \cite{SWi}. It is very
important to appreciate the difference between classical and quantum
localization. For the FWG fields the quantum localization is defined by
these intersection of algebras and cannot be replaces by the classical
localization in the sense of support properties of test functions. 

It is instructive to reproduce the results of Smirnov and the reformulation
of Laskevich \cite{Smi}\cite{Lash} in the present setting.

\begin{theorem}
A sufficient condition for a power series in the $Z$-fields to describe a
pointlike local field $A(x)$ is in addition the previous commutation (\ref
{co})and cyclicity property (\ref{cyc}) (in case of our illustrative model
without bound states) the presence of poles in the $a_{n}(\theta _{1},\theta
_{2}....\theta _{n})$ for $\theta $-differences lying on the boundary of the
strip: 
\begin{equation}
a_{n+2}(\vartheta +i\pi +i\varepsilon ,\vartheta ,\theta _{1},\theta
_{2}....\theta _{n})\simeq \frac{1}{\varepsilon }\left[
1-\prod_{i=1}^{n}S(\vartheta -\theta _{i})\right] a_{n}(\theta _{1},\theta
_{2}....\theta _{n})
\end{equation}
\end{theorem}

Here we did not specify the Lorentz transformation property of the field 
\begin{eqnarray}
A(x) &=&\sum_{n}\frac{1}{n!}\int_{C}d\theta _{1}...\int_{C}d\theta
_{n}e^{-iP(\theta _{1},...\theta _{n})x}a_{n}(\theta _{n},\theta
_{n-1}....\theta _{1}):Z(\theta _{1})....Z(\theta _{n}):  \nonumber \\
&&P=\sum p_{i}(\theta )
\end{eqnarray}
If the field is irreducible with spin s one must have 
\begin{equation}
a_{n}(\theta _{1}+\vartheta ,\theta _{2}+\vartheta ....\theta _{n}+\vartheta
)=e_{n}^{-s\vartheta }(\theta _{1},\theta _{2}....\theta _{n}),
\end{equation}
but in the proof this is not needed. The proof is purely structural i.e.
independent of the computation (or parametrization of the solutions of all
these conditions).

\textbf{Proof }

we show that for spacelike separation we can by contour deformation
transform of 
\begin{eqnarray}
&&A(x)B(y)=\sum \frac{1}{m!n!k!}\int_{-\infty }^{\infty }d^{k}\xi
\int_{C}d^{m}\theta \int d^{n}\vartheta \times  \\
&&\times \exp -i\left[ P(\xi )(x-y)+iP(\theta )x-iP(\vartheta )y\right]  
\nonumber \\
&&\times a_{m+k}(\xi _{1},...,\xi _{k},\theta _{1},...,\theta
_{m})b_{n+k}(\vartheta _{n},...,\vartheta _{1},\xi _{k}-i\pi ,,...,\xi
_{1}-i\pi )  \nonumber \\
&&\,\,\,\,\,\,\,\,\,\,\,\,\,\,\,\,\,\,\,\,\,\,\,\,\,\,\,\,\,\,\,\,\,\,\,\,\,%
\,\,\,\,\,\,\,\,\,\,\,\,\,\,\,\,\,\,\,\,\,\,\,\,\,\,\,\times :Z(\theta
_{m})...Z(\theta _{1})Z(\vartheta _{1})...Z(\vartheta _{n}):  \nonumber
\end{eqnarray}
achieve the opposite order. Since the $Z^{\prime }s$ underneath the
Wick-product do not commute (unlike the previous case of $Z^{J}$ with $Z),$
the compensating terms must come from the $i\pi $ shift of contour, i.e.
from residua of poles encircled in the process of deformation. In order to
apply (\ref{form}) say to $\xi _{i}$ together with $\theta _{j}$ (i.e. the
pole at $\xi _{i}=\theta _{j}+i\pi )$, we have to position the relevant
variables at the first two places in $a_{m+k}$ by successive transpositions.
Whereas the contributions from $\xi $-permutations in $a_{m+k}$ compensate
those in $a_{n+k},$ from the $\theta $ one obtains the phase factor $\Pi
_{l\neq i}S(\xi _{l}+i\pi -\theta _{j}).$ It is convenient to shift
simultaneously $Z(\theta _{j})$ to the right and denote the total resulting
phase factor: 
\begin{equation}
s(\xi _{i})\equiv \Pi _{l\neq i}S(\xi _{l}+i\pi -\theta _{j})\Pi
_{l=1}^{n}S(\theta _{j}-\vartheta _{l})  \label{s}
\end{equation}
Now we do the contour shifting which according to (\ref{form}). We only pick
up poles from the coefficient function $a_{m+k}$. As a result we get the
additional factor -$\left[ 1-\Pi _{l\neq i}S(\theta _{j}-\xi _{l})\Pi
_{l\neq j}S(\theta j-\theta _{l})\right] .$ Note that for space-like
distances (we can place the x,y so that they are in opposite wedges) the
exponential factor is damped in $0<Im\xi <\pi .$ The integrand of the term $%
(m,n,k)$ is, apart from the mentioned two factors and the exponential
function, as follows 
\begin{eqnarray}
&&a_{m+k-2}(\xi _{1},..,\hat{\xi}_{i},..,\xi _{k},\theta _{1},..,\hat{\theta}%
_{j},..,\theta _{m})b_{n+k}(\vartheta _{n},..,\vartheta _{1},\xi _{k}-i\pi
,,..,\theta _{j},..,\xi _{1}-i\pi ,\theta _{j})  \nonumber \\
&&\,\,\,\,\,\,\,\,\,\,\,\,\,\,\,\,\,\,\,\,\,\,\,\,\,\,\,\,\,\,\,\,\times
:Z(\theta _{m})...\stackrel{\symbol{94}}{Z(\theta _{j})}...Z(\theta
_{1})Z(\vartheta _{1})...Z(\vartheta _{n})Z(\theta _{j}):
\end{eqnarray}
where as usual the roof sign indicates deletion of the variable or the
operator. commuting the operator $Z(\theta _{j})$ to the left, we obtain a
phase factor which cancels the previous factor evaluated at the pole $s(\xi
_{i}=\theta _{j}+i\pi ).$ The important step is now the following renaming
(leading to a resummation in $\sum )$: $\theta _{j}\rightarrow \vartheta
_{n+1},$ followed by $n\rightarrow n-1,m\rightarrow m+1$ and $k\rightarrow
k+1$ and re-numbering $\xi $ and $\theta $%
\begin{eqnarray}
&&-\left[ 1-\Pi _{l\neq i}S(\theta _{j}-\xi _{l})\Pi _{l\neq j}S(\theta
j-\theta _{l})\right] a_{m+k}(\xi _{1},..\xi _{k},\theta _{1},..,\theta _{m})
\\
&&\,\times b_{n+k}(\vartheta _{n},..\vartheta _{1},\xi _{k}-i\pi ,..\xi
_{1}-i\pi ):Z(\theta _{m})..Z(\theta _{j})..Z(\theta _{1})Z(\vartheta
_{1})..Z(\vartheta _{n}):  \nonumber
\end{eqnarray}
This shifting process has to be repeated with every $\xi $ and the boundary
term has to be taken into account (i.e. the remaining integration variables
are placed on the boundary $\xi \rightarrow \xi +i\pi ).$ Using the
following identity 
\begin{equation}
\sum_{l=0}^{n}\left( -\right)
^{l}\sum_{j_{1}<...j_{l}}\prod_{r=1}^{l}(1-t_{j_{r}})=\prod_{j=1}^{n}t_{j}
\end{equation}
one arrives at the desired result 
\begin{equation}
A(x)B(y)=B(y)A(x),\,\,\,\left( x-y\right) ^{2}<0
\end{equation}

As mentioned before this derivation is not quite in the spirit of our
modular approach. The latter requires to study the double cone algebras i.e.
to characterize the quantum localization properties of intersections and to
derive a necessary and sufficient condition on the $a_{n}$ \cite{SWi}$.$

One can show that the modular construction of ``free'' anyons and plektons 
\cite{schroer} in d=1+2 leads to similar mathematical problems. In this case
there is no scattering, but the whole construction takes place in a
multiparticle space which, in contradistinction to Fermions and Bosons, has
no tensor product structure in terms of Wigner spaces. The braid group
commutation relation leads to a Tomita $J$ which, as in the Federbush model,
involves a constant matrix $S_{twist}$. In this case $S_{twist}$ does not
carry scattering information, but is identical to the braid group
representation R-matrix whichwhich appears in the exchange algebra of
conformal QFT. This statistical R-matrix $S_{twist}$ has a similar effect as
a Klein transformation i.e. the opposite of $\mathcal{A}(W)$ in the quantum
sense of the von Neumann commutant is now \textit{different from the
geometric opposite} $A(W)^{\prime }\neq A(W^{\prime }).$ Again the model has
a rich virtual particle structure, even though no real particle is created
in a scattering process. But since this time this vacuum polarization is a
result of the nontrivial statistics twist $S_{twist},$ there is no reason to
expect that this goes away in the nonrelativistic limit and the model passes
to Schr\"{o}dinger theory. The fact that the nonrelativistic limit of
Fermions and Bosons leads to the Schr\"{o}dinger QM is related to the 
\textit{existence of relativistic free fields in Fock space}. But since the
Fock space structure in d=1+2 cannot support anyons and plektons, there is
good reason to expect a kind of \textit{nonrelativistic field theory which
can incorporate the virtual particle or vacuum polarization structure} which
is \textit{necessary} to \textit{maintain the relation between spin and
(plektonic) statistics in the nonrelativistic limit}. Indeed all attempts to
incorporate braid group statistics into QM, ever since the time of Leinaas
and Myrheim \cite{LM} \cite{Mund Schrader} have only led to a deformation of
(half)integer spin \cite{Wilczek}, but not to nonrelativistic operators with
the correct spin-statistics commutation structure. To phrase it into a more
mundane fashion: ther are good reasons why nobody has succeeded to construct
a QM interaction (e.g. Aharonov-Bohm like) which leads to an anyonic spin in
the two-particle S-matrix and fulfills those higher particle S-matrix
cluster properties which are necessary in order to obtain coherence with the
multiparticle statistics. The attempts based on Aharonov-Bohm potentials
allow only to mimic an anyonic two-particle spin. I would not  expect that
the scattering boundary condition for the long range A-B interaction of
n-particles can be chosen in such a way that the multiparticle S-matrix
fulfills the cluster property which certainly would be a prerequisite for
sustaining a nonrelativistic spin-statistics connection. Like e.g. the
nonrelativistic Lee model, I rather expect anyons and plektons to be only
describable by a \textit{nonrelativistic field theory} which maintains the
vacuum polarization and this is my explanation for why the search for a
consistent theory of particles with genuine braid group statistics in
quantum mechanics has been unsuccessful.. At this point it is instructive to
compare with a recent No-Go theorem \cite{Mund}.

The method advocated here by analogy to factorizing d=1+1 models of using
the Wigner one particle representations combined with the correct
non-tensor-product multiparticle structure from scattering theory \cite{FGR}
together with the present modular localization method looks very promising
(but still needs to be carried out). The results should be of great
practical relevance for unraveling the structure of (nonrelativistic) d=1+2
quasi-particles in many body systems with spatial layer structures \cite
{Laughlin}. As with Fermions and Bosons, the derivation of the
spin-statistics properties and the classification of the possible statistics
(+internal symmetry) is done in the relativistic setting, in order then to
be used e.g. for the (nonrelativistic) quasi-particles in condensed matter
physics and statistical mechanics.

A similar idea of using an $S_{twist}$ in a constructive modular approach
should also be helpful to complete the classification and construction of
chiral conformal QFT's. In this connection one should recall that presently
the construction of these models is done in most cases by studying the
representation theory of affine algebras, in other words by ideas which are
quite different from those of standard QFT. The charge-carrying fields which
fulfill braid group exchange algebras are then constructed as intertwiners
between the vacuum and charged representations of these algebras. It would
be more in the spirit of QFT to first classify all plektonic statistics (the
structure constants of the exchange algebras) and then to construct the
vacuum representation (expected to be unique) of the associated exchange
algebras. There is however one caveat: the \textit{exchange algebra} (unlike
the CCR or CAR algebras) \textit{is incomplete} since the distributional
behavior at coalescent points is left undetermined. This is the reason why
the attempts in e.g. \cite{S-R} which were based on monodromy properties are
more ``artistic'' (in the sense of depending on hindsight and luck) than
systematic. Momentum space algebras, as the Z-F algebra on the other hand,
are complete. But can one use the R-matrices as a Klein twist $S_{twist}$ in
momentum space using the previous method of wedge localization? Since
conformal theories have an infraparticle- but not a particle-structure, this
problem is not trivial. The normalization for obtaining finite form factors
is infrared divergent with respect to that of finite correlation functions.
We hope to present an affirmative answer in a separate work. The importance
of such a new approach to chiral conformal QFT results from the fact that it
places it back into the mainstream of QFT where it belongs. The statistics
input in the absence of genuine interaction (chiral conformal QFT's are
expected to belong to this kind since there can be no genuine coupling
parameter dependent interaction on one light cone) and for given internal
symmetry groups (in order to have a distinction between charged fields for
current algebras and the W-fields) should uniquely determine the vacuum
representation of the completed exchange algebra. In more technical terms
and only for experts on conformal QFT: the Friedan-Qiu-Shenker-quantization
of the energy-momentum tensor algebra and similar quantizations for
W-algebra generalizations should follow from the more fundamental
DHR-Jones-Wenzl quantization which has a direct interpretation in terms of
particle statistics. In this way chiral conformal QFT would change from a
mathematical playground back to a valuable theoretical laboratory of QFT
since it can then be formulated as simple analytic realization of principles
which have validity in higher dimension.

In the remaining part of this section I would like to make some pedagogical
remarks for readers with an incomplete knowledge of general structural
properties of nonperturbative QFT. In connection with the analytic aspects
of the rapidity-dependent Zamolodchikov-Faddeev algebra operators and their
analogy with the x-space chiral conformal operators of the exchange algebras
one often finds the erroneous concept of ``analytic field operators'' and
``holomorphic algebras''.$\,$Since their use is so widespread (the few
articles where this misleading terminology is not used are rather the
exception than the rule), it is interesting to ask where such ideas are
coming from. I am not an expert on string theory, therefore I have limited
my search to QFT. The oldest paper which could be interpreted as alluding to
``analytic operators'' $A(z),z\in \mathbf{C}$ seems to be the famous BPZ 
\cite{BPZ} paper\footnote{%
In an older paper on conformal blocks e.g. \cite{SSV} (called nonlocal
components in a conformal decomposition theory with respect to the center of
the universal covering), such ``holomorphic'' terminology was never used.
\par
.} on minimal models. Although the authors do not use such terminology in
print, the notation used in that paper may have caused misunderstandings
(and has been misunderstood by physicist whose first experience with
nonperturbative QFT came through that famous work or was influenced by
string theory). The truth is that \textit{field algebras never have
holomorphic properties}. The analytic properties of correlation functions
and state vectors depend entirely on the nature of states one puts on those
algebras. Whereas vacuum ground states lead to the famous BHW-domain \cite
{Wig Str} (in chiral conformal QFT equal to a uniformormization region with
poles for coalescing coordinates), KMS states will only lead to strip
analyticity. It is in any case the state which generates the analytic
continuation holomorphy properties, and the higher its symmetry property,
the bigger the holomorphic regions.

It is a fact that the associated analytic Bargman-Hall-Wightman domain for a
Moebius invariant vacuum state in conformally covariant QFT is larger than
that of the corresponding Poincar\'{e} covariant massive theories and as a
result of braid group statistics, one looses the univaluedness of these
analytic continuations (but of course not the univaluedness in the real time
physical localization points!). The effect of restriction of the vacuum
state of massive theories to the wedge algebra yields the momentum space
analytic properties which together with the factorization property lead to
the rapidity being an analytic uniformization variable. Contrary to the use
of field theoretic terminology in the contemporary literature neither the
old nor the new BHW domains nor the analytic continuation in rapidity have
anything to do with the ``living space'' of fields in the sense of quantum
localization of operators in this article. ``Localization'' is indeed a
property of the algebras whereas holomorphy is not.

In an attempt to attribute despite the negative previous remarks a
constructive algebraic meaning to the above unfortunate but nevertheless
popular terminology, one could point to the following property which has 
\textit{no analogue in higher dimension} (not even in the conformal
invariant limit of higher dimensional theories). Whereas generally with
vacuum expectation values one can relate at most two physical theories: a
noncommutative real time QFT and a commutative euclidean field theory (a
candidate for a continuous statistical mechanics), a chiral conformal theory
on one light cone has \textit{infinitely many noncommutative boundary values}
(and no commutative chiral boundary value) each of which defines a set of
positive definite correlation functions and hence a theory. This is to say
the restriction of the analytically continued correlation function defines a
positive QFT not only on the circle (the standard living space of chiral
theories) but also \textit{on each boundary encircling the origin} (with the
right $i\varepsilon $ Wightman boundary prescription)\footnote{%
This is probably what string theorist have in mind when they draw their
pictures.}. However all this does not legitimize ``holomorphic operators''
in the literal sense but rather the existence of a operator conformal QFT
for each chosen boundary circumference. The reader recognizes easily that
this structure is equivalent to the existence of the infinite dimensional
diffeomorphism group which is related to the Virasoro algebra structure. The
application of any symmetry, which does not leave the vacuum reference state
invariant, defines another set of positive Wightman functions which at the
case at hand belong to the deformed boundary. If one prefers a geometrical
to a quantum physical terminology, one may emphasize the diffeomorphism
group, whereas for a physicist who prefers the setting of local quantum
physics the existence of a an interesting generalization of the higher
dimensional dichotomy between real time-imaginary time theories may be the
interesting aspect.

\section{ General Interactions and Outlook}

The factorizing models of the bootstrap formfactor approach belong to a
class of models which are ``real particle'' (on shell) conserving but
``virtual particle'' (off shell) nonconserving. In the context of our
modular localization approach this means that although the wedge
localization Hilbert spaces can be generated by on shell FWG operators, any
sharper localization as e.g. the double cone localization and in particular
the state vector obtained by applying smeared pointlike local fields to the
vacuum lead to ``virtual particle clouds''. Although absence of real
particle creation according to well-known theorems is impossible in d=1+3
interacting theories, we expect the three-dimensional theories d=1+2 to form
an exception if the associated particles obey genuine braid group statistics
(anyons and plektons) and hence their charge-carrying field have a
noncompact semiinfinite string-like extension \cite{schroer}. Namely we
expect the existence of ``free''\footnote{%
If ''free'' in d=1+2 implies the existence of generating fields $A$ in Fock
space without vacuum polarization clouds in $A\Omega $ then plektons are
never free because the vacuum polarization is needed to sustain the braid
group statistics and the associated string-like extension. We use this
terminology in the more liberal sense of absense of additional interactions
which cause nonvanishing scattering cross sections.} anyons and plektons
which similar to free Fermions and Bosons have no on-shell creation but
(different from free Fermions and Bosons) possess a rich virtual particle
structure \cite{Mund}.

Let me finally address the important question of whether the concept of
modular localization can be expected to lead to a nonperturbative approach
for d=1+3 interacting theories of Fermions and Bosons. This depends on
whether it is possible to relate an admissable S-matrix with an auxiliary
semilocal (wedge localized) operator $Z$ which is free of vacuum
polarization and creates the interacting localized wedge spaces $\mathcal{H}%
_{R}(W).$ The existence of such an ``on-shell'' operator would be the
correct substitute for the FWG operators in the case of non-factorizing
situations. On a formal level such operators have appeared in the light cone
quantization of interacting theories \cite{Susskind} but in that formalism
the relation to the original local variables (with vacuum polarization) gets
lost and without relation to local operators one has no physical
interpretation beyond global spectral properties \cite{Notes}. Our modular
wedge localization approach on the other hand does not depend on assumptions
about canonical commutation relations and therefore could serve as a
rigorous substitute for the ill-defined light cone quantization.

A prerequisite for all modular constructions including the modular approach
to the formfactor bootstrap program is its uniqueness i.e. to one admissable
S-matrix there should be only one local algebraic net. This is part of the
more general question whether modular data ($J,\Delta ^{it})$ determine
uniquely an algebra and a state. Whereas this modular inverse problem has
many solutions \cite{Wollenberg}, the corresponding inverse problem of QFT 
\begin{equation}
S_{sc}\,\,in\,\,\mathcal{H}_{Fock}\stackrel{?}{\rightarrow }\left\{
A(W)\right\} _{all\,\,wedges}
\end{equation}
can be shown to have a unique solution even beyond the special factorizing
d=1+1 models \cite{SWi} if one accepts certain plausible but presently
unproven) vacuum and one-particle properties of the modular M\o ller
operator $U$ \cite{SWi}$.$

Acknowledgments:

I am indebted to Hrach Babujian, Andreas Fring and Michael Karowski for many
discussions. I owe special thanks to Michael Karowski for his patience with
which he explained many subtle points about the structure of formfactors as
well as to Karl-Henning Rehren for a helpful correspondence.. Finally I am
thankful to Hans-Werner Wiesbrock for reading the manuscript and in
particular for his active interest in the use of modular ideas for the
construction of interacting theories.

\end{document}